# ABOUT A MECHANISM OF THE SHORT PERIOD 160-MIN. RADIAL PULSATIONS OF SUN

*Ivan T. Ivanov[(1)], Valentina T. Taneva[(2)] and Boris Komitov[(3)]*

(1) Department of Physics and Biophysics, Medical Institute, Thracian University, 6000 Stara Zagora;

(2) Institute of Applied Physics, Technical University of Sofia

(3) Bulgarian Academy of Sciences - Institute of Astronomy, 6003, Stara Zagora-3, P.O. Box 39
b_komitov@sz.inetg.bg

The visible diameter of Sun oscilates with a period of 160 min. The same type of periodicity is also found in a huge number of solar radiation parameters. To elucidate the origin of these longitudinal radial pulsations we have used the equation for the equilibrium of inner layers which, after linearization, turned into the harmonic oscilator equation. The latter equation allows radial pulsations whose period and wave length were calculated using regresion expresions for the gas presure and density in various layers. The radial pulsations originate at the surface of active zone and propagate til the litosphere, where they undergo full inner reflection producing undersurface stationary waves with a period of 150-160 min.

INTRODUCTION.

Ever since the discovery of the 160-min radial pulsations of the Sun ample data have been collected suggesting their impact on the sollar physics [1,2]. 160-min oscilations are observed in the intensity of light, radio and infrared radiation emitted from the resting regions of Sun [3]. The pulsations observed in the radiobrightness lag behind the cyclic radial movements by about 12 min [3]. A 0.02% variation in the intensity of the 1.65 mm infrared sollar radiation is detected with a 160-min period [4]. Oscilations with the same period are established in the cyrcular polarized radio emition of Sun at 13.5 mm, which lag behind the oscilations of the radiant velocity by 34 minutes [5].

These observations could not be explained as pure atmospheric effects [6]. The radiant pulsations of Sun have been unsuccefully related to the particularities of the radiant energy transfer [7], to the possible interference between the Sun gravity modes [8] or even to the possible existence of a particular object orbiting the Sun center at appr. 20 000 km under the surface of the Sun [9]. Thus at present there is no commonly accepted theory that could explain the radiant pulsations of the Sun. In this paper we offer a model based on the linearization of the Sun status equations. This model reprisents the radiant oscilations as staying waves generated on the surface of the active zone and propagating to the surface of the Sun where they have a period of about 160 minutes.

**THE MODEL.**

The equilibrium status of the star core is usually found as a solution of the system of basic equations describing the star physics. For the Sun, several solutions are obtained which give the equlibrium values of pressure *P*, temperature *T* etc as functions of the distance *r* from solar center. These functions are averaged in the summarised model [10] which will be used futher.

The status of the Sun intestine is a stable equilibrium. If a thin concentric layer inside the Sun is displaced from its equilibrium distance $r_0$ from the center of gravity, it will commence to oscilate. What will be the frequency of these oscilations? The equation that describes the movement of the layer is

$$-\rho \cdot (\partial^2 r / \partial t^2) = \partial P / \partial r + \frac{\rho \cdot G \cdot m}{r^2}$$

(1)

where *m* is the mass of the gas placed under the layer. Let this layer be shifted from a place with distance $r_0$ to a place with a distance $r_0 + \Delta r$. Since $\Delta r \ll r_0$, the pressure gradient will change as given by

$$\partial P / \partial r = (\partial P / \partial r)_{r=r_0} + (\partial^2 P / \partial r^2)_{r=r_0} \cdot \Delta r + o(\Delta r^2)$$

(2)

The gravity force attracting the layer to the center of Sun will also change according to the variations in the distance *r*

$$\frac{1}{r^2} = \frac{(1 - 2 \cdot \Delta r / r_0)}{r_0^2} + o[\Delta r / r_0)^2]$$

(3)

Let us combine the equations (2) and (3) with the equation (1). Inserting a new variable $x = \Delta r / r_0$, a new equation will be obtained

$$\partial^2 x / \partial t^2 + \frac{1}{\rho_0}(\partial^2 P / \partial r^2) - 2Gmr_0^3)x = \frac{\rho_0}{r_0}(\frac{\partial P}{\partial r}) - \frac{Gm}{r_0^3}$$

(4)

The right side of the latter equation is equal to zero since a hydrostatic equilibrium holds at $r = r_0$. Equalizing the left side of the equation to zero a new equation will be obtained which resembles the harmonic oscilator relation. Thus, the respective frequency f of the adiabatic radiant oscilations will be given by the formula

$$f^2 = \frac{1}{\rho}(\partial^2 P / \partial r^2) - \frac{2Gm}{r^3}$$

(5)

The period of oscilation of the layer will be $\Pi = 2\pi / f$. Once generated these oscilations will apparently subside down provided there is no source of energy for their rejuvanation.

The above linearizations appear fairly correct since the visible diameter of the Sun is found to oscilate with an amplitude of only a few km and a radiant velocity of about several m/s. Similar pulsations are found for the variable stars however the corresponding parameters of pulsations are three orders of magnitude greater in respect to these of Sun. Nevertheless, similar linearization theory has been also applied for the variable stars giving constant amplitude, frequency and phase at different distances from the star center [11]. In contrast to the oscilations of variable stars, the oscilations that could occur within the Sun are very small and should be consequently allowed to have different amplitude, frequency and phase at different distances from the center of Sun as given by the formula (5).

RESULTS

According to formula (5), the period $\Pi$ of oscilation of different layers could be calculated using proper expresions for the dependences of P, r and m on the relative distance $x = r/R$ from the Sun center. Using the model [10] and 16 cited values for each parameter, these expressions were obtained as regression formulae and are shown in Table I. For each parameter the correlation coefficient $Kr$, reliability factor $F$ and probability weight factor $Pw$ between the cited data and the calculated values demonstrate that the obtained expressions are fairly satisfactory.

*Table I. Regression formulae for the density $\rho$ and gas presure P within the Sun as functions of the relative distance $y = r/R$ from its center. m is the mass of solar gas beneath a concentric layer with a radius r. R and M are the Sun radius and mass correspondingly.*

| Regression formula | Kr | F | Pw |
|---|---|---|---|
| m = M. (2.84 y - 1.70 y - 0.06) | 0.991 | 52.1 | 4.10 |
| P = exp (38.94 - 24.68 y ) | 0.920 | 6.08 | 0.0008 |
| R = exp ( 3.85 -17.36 y ) | 0.942 | 8.28 | 0.00015 |

Allowing x to vary with a step of $\Delta x = 0.01$, a set of discrete values of $\Pi$ were calculated which are shown on plot (Fig.1). Generally, the intestine of Sun is sharply divided into three zones; active core, intermidian zone and convective zone, according to the nature of the physical processes in them. Each of these zones is clearly distinguished on the given plot.
Oscilations should be apparently imposible throughout the active zone ($0 < x < 0.38$) as $f^2 < 0$. It could be assumed that the period of oscilations there will be close to infinity. In the vast convective zone of radiative energy transfer, oscilations might occur with a nearly constant period of about 1000 s. However, at the boundary between this zone and the active zone the period of pulsations sharply increased inclining to infinity (Fig.1) in accordance with the result that ascilations are not allowed in the active core. This result indicates that any slow mechanical disturbance that might occur on the surface of the active zone could be transfered into the interior of the middle zone as periodic pulsations. Hence, the radial oscilations that might occur within the Sun could originate from this boundary. Such a conclusion is apprehensible since the surface of the active zone must be mechanically unstable.

According to Fig.1, the period of pulsations slowly increases near the outside boundary of the convective zone reaching the the value of 8900 s (148 min) just on the Sun surface. This value differs by about 8 % from the experimentally obtained value of

160 min. The striking coincidence between the calculated and measured values of solar pulsations supports the proposed model.

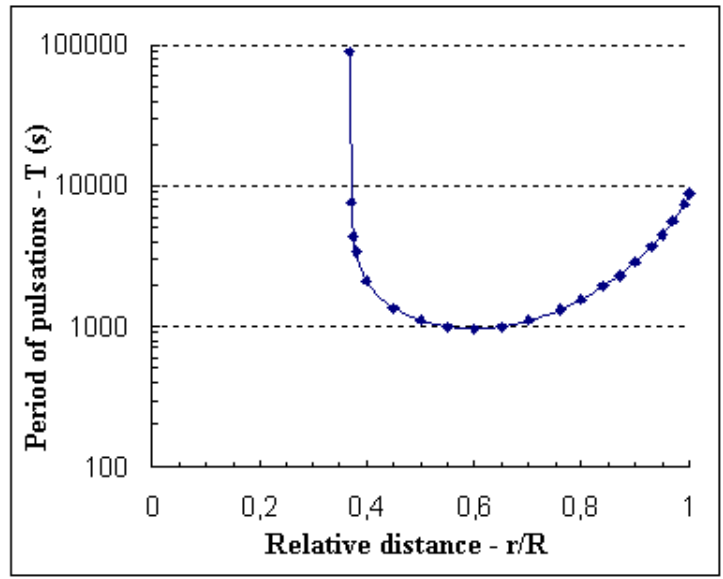

*Fig.1. Period of radial pulsations within the Sun as a function of the distance from its center.*

DISCUSSION

According to the proposed model, the pulsations of the Sun radius are brought about by a slight disturbance of the dynamic equilibrium of forces balancing the inner layers and could be decsribed by the mechanism of the charmonic oscilator. The energy source generating the oscilations and maintaining their amplitude constant is possibly found at the boundary between the active and the middle zones. Once generated, the oscilations should further propagate towards the Sun surface with the velocity "$C$" of the sound. This velocity could be calculated using the formula $C = (P/\rho)^{1/2}$ and the expressions for the P, and p as given in Table I. We have calculated the velocity C and the wave length $\lambda = \Pi.C$ of these mechanical pulsations as a function of the distance $x$ form the center of Sun. Suprisingly $\lambda$ was nearly constant for both the middle and convective zones (data not shown).

When the radial oscilations reach the Solar atmosphere (the chromosphere), they should sustain a complete backward reflection without change in phase as suggested by the strong fulfilment of the respective mandatory condition $\lambda > 4\pi H$. Here $H$ is the depth of the solar atmosphere. As the wave length is practically constant within a broad segment under the visible surface of Sun, the formation of staying mechanical wave shoud be allowed deep under the chromosphere. These oscillations have the capacity to impact a huge number of physical phenomena close to the surface of Sun introducing a variable component with the same 160 min periodicity in the observed parameters of Sun.